\documentclass[12pt,preprint]{aastex}
\usepackage{amssymb,amsmath,float}
\newcommand\bea{\begin{eqnarray}}
\newcommand\eea{\end{eqnarray}}
\newcommand\fr{\frac}
\newcommand\Ph{\Phi}

\newcommand\tha{\theta}
\newcommand\ta{\tau}

\newcommand\om{\omega}
\newcommand\rb{{\tilde r}}
\newcommand\br{{\tilde r}}
\newcommand\lpar{\left(}
\newcommand\rpar{\right)}

\begin{document}
\title{The outermost bound orbit 
around a mass clump in an
expanding Universe: implication on rotation curves and
dark matter halo sizes}

\author{Richard Lieu$\,^{1}$}

\affil{\(^{\scriptstyle 1} \){Department of Physics, University of Alabama,
Huntsville, AL 35899.}\\}

\begin{abstract}

Conventional
treatment of cold dark matter halos employs the Navarro-Frenk-White (NFW)
profile with a maximum radius set at $r=r_{200}$, where the enclosed matter
has an overdensity of 200 times the critical density.
The choice of $r=r_{200}$ is somewhat
arbitrary.  It is not the collapsed (virial) radius, but does 
give $r \sim$ 1 Mpc for rich
clusters, which is a typical X-ray size.
Weak lensing measurements, however, reveal halo radii well in excess of 
$r_{200}$.  Is there a surface that places an absolute limit 
on the extension of a halo?
To answer the question, 
we derived analytically the solution for circular orbits
around a mass concentration in an expanding flat Universe, to show that
an outermost orbit exists at $v/r = H$, where $v$ is the orbital speed
and $H$ is the Hubble constant.  The solution, parametrized as $r_2$,
is independent of model assumptions on structure formation, and
{\it is the radius at which the furthest particle can be regarded as
part of the bound system}.
We present
observational evidence in support of dark matter halos reaching 
at least as far out
as $r=r_2$.  An interesting consequence that emerges concerns the
behavior of rotation curves.  Near $r=r_2$ velocities will be biased low.
As a result, the mass of many galaxy groups may have been underestimated.
At $r=r_2$ there is an
abrupt cutoff in the curve, irrespective of the halo profile.
An important cosmological test can therefore be performed if velocity disperion
data are available out to 10 Mpc radii for nearby clusters (less at higher
redshifts).  For Virgo it appears that there is no such cutoff.

\end{abstract}

\noindent
{\bf 1. Introduction}

In the theory of large scale structure formation, a pillar of the
standard cosmological model,
one important question that has not been answered to satisfaction
concerns where exactly, viz., at what threshold of overdensity, is the
boundary of a mass concentration (or clump) at a given redshift.
The question is pertinent
to any effort in obtaining a reliable theoretical estimate of
the total mass and spatial extent of the `dark halos'
in galaxies and clusters.  The reason has to do with the availability, from
numerical hydrodynamic simulations (Navarro, Frenk, and White 1995, 1996,
1997;  but see also the even earlier work of Dubinski \& Carlberg 1991), of
a `universal' profile (known as NFW profile) for the clumps, of the form
\begin{equation}
\rho_m (\br) = \frac{\delta_c \rho_c r_s^3}{\br(\br +r_s)^2}
\end{equation}
for the matter density distribution, where in
Eq. (1) $\br$ denotes a {\it physical} radius (i.e. an invariant for
bound structures),
$r_s$ is a constant scale radius, $\delta_c$ is an overdensity
parameter, and $\rho_c$ is the critical density.
Since at the outer
radii $\br \gg r_s$ the density scales as $\rho_m (\br) \sim 1/\br^3$,
{\it and there is no further change of functional form with increasing
distance}, i.e. hierarchical structure formation codes do not seem
to reveal the surface radius of a clump,
the total integrated mass is divergent unless an upper radius limit,
(or cutoff) $R$ , is `manually' assigned.  

Conventionally this
limit is set at $R=r_{200}$, sometimes referred also to
as the virial radius, defined as the radius at which the enclosed matter
density is 200 times above the critical density
\begin{equation}
\rho_c = \frac{3 H_0^2}{8\pi G}
\end{equation}
The choice of the virial radius $R$ is probably just a matter
of convenience, e.g. for a rich cluster like Coma one
may envisage a virial mass $\sim$ 2 $\times$ 10$^{15}$ M$_\odot$,
in which case a factor of 200 overdensity would lead,
via the equation
\begin{equation}
M_{200} = \frac{800\pi}{3} \rho_c r_{200}^3.
\end{equation}
to $R \approx$ 2.5 Mpc, which is not far from the value
believed to be `reasonable'.
(Lokas \& Mamon 2003).  Historically, an analytical formula for the virial
radius was derived by directly appealing to the virial theorem (Lahav
et al 1991).  For an
$\Omega_m =$ 0.27 and $\Omega_{\Lambda} =$ 0.73 cosmology 
(e.g. Spergel et al 2006)
this radius (sometimes also called the
`collapsed' radius) is less than half the `turnaround' 
radius of Peebles (1984),
and corresponds to an overdensity of $\approx$ 90$(1+z)^3$:
see Fig. 1 of Eke,
Cole, and Frenk (1996).
As we shall see, the actual evidence points to a continuation
of the NFW profile to radii well beyond this value, suggesting the
already well accepted fact that
particles need not be completely virialized before they become part of
the clumped (or collapsed) system.  

If the virial radius lacks observational
significance
are there other criteria available that may better connect
with reality?
The fundamental point here is that for a pure
Newtonian (static and infinitely old) Universe there is in principle no
end to the zone of gravitational influence of a clump.  Clearly the
same is no longer true for an expanding Universe.  Can an outermost surface
for a bound structure be drawn, and be subject to scrutiny?
Attempts to define such a surface
based upon consideration of radial motion, 
have been made (e.g. Sandage 1986).  The
question which cannot be answered by simply calculating radial
velocities, however, is whether
bound orbits can still exist at these large distances.
As it turns out, there {\it is} a radius within
which circular orbits are sustainable, 
The derivation of this radius does not depend
on any of the assumptions made about how the clump is formed.

\vspace{2mm}

\noindent
{\bf 2. Criterion for the existence of bound orbits in an
expanding Universe}

If in a flat and unperturbed FRW space-time with dimensionless
expansion parameter $a(t)$ and $c=1$, i.e.
\begin{equation}
ds^2 = dt^2 - a^2 (t) (dr^2 + r^2 d\Omega^2)
\end{equation}
we envisage particles located
at constant physical distances from each other, or geometrical shapes
that do not expand with the Universe, it would be more convenient to
use $\rb = ar$ rather than $r$ as radial coordinate, because our objects of 
interest are tied to the $ar$-grid and not the $r$-grid.  Moreover, if we also
perform the transformation $t \rightarrow \tilde t$ 
where $t = \tilde t - H \tilde r^2/2$, Eq. (4) will become
Minkowski in form,
\begin{equation}
ds^2 = d\tilde t^2 - (1+H^2 \tilde r^2)
(d\tilde r^2 -
\tilde r^2 d\Omega^2),
\end{equation}
apart from one correction term which is only second order in $H$ (we assumed
zero acceleration of the expansion by ignoring a $qH^2$ term; as will be
evident from the complete treatment below, the effect of $q$ is negligible
for our purpose).

Eq. (5) is the reason
why no experiments  performed on earth, or within the Milky Way, can
directly probe the Hubble expansion.  An interesting
point emerges nonetheless when one computes the radial speed of light
$$
\frac{d\tilde r}{d\tilde t} = 1 - \frac{1}{2}H^2 \tilde r^2,
$$
and find that to order $H^2$ it decreases with increasing $\tilde r$,
symptomatic of 
an effective
repulsive potential that eventually prevails at large radii
where expansion might overcome gravity.  At this stage, however,
the idea is only suggestive, because we have yet to formally include
the effect of gravity.

If the large scale character of space-time is that of a flat FRW  metric,
but locally there is perturbation by
a weak and centrally symmetric gravitational field, the line element will
be modified to the form
\begin{equation}
ds^2 = (1+2\Phi) dt^2 - (1-2\Phi) a^2(t) (dr^2 + r^2 d\Omega^2),
\end{equation}
where $\Phi$ is the gravitational potential
(McVittie 1933, see also
Futumase \& Sasaki 1989).  This time the transformation to coordinates
$(\tilde t, \tilde r)$  no longer offers  a great deal
of simplification.  Nevertheless,
McVittie (1933) showed that circular orbits maintain their constant
radii $\rb = a(t)r$,  so there is still some mathematical
advantage to be gained in
preferring $\rb$ to
$r$ as the radial coordinate for our problem, i.e. in a $(t, \rb)$ system 
Eq. (7) reads as
 \begin{equation} ds^2=[1+2\Ph(\rb)]dt^2-[1-2\Ph(\rb)]
 [(d\rb-H\rb dt)^2+\rb^2d\Omega^2]. \end{equation}
Although
in Eq. (8) one encounters a cross term of the form $H \tilde r d\tilde r dt$ 
which
is first order in the Hubble constant, we shall soon find out when we
calculate orbital behavior
that, in agreement with our earlier analysis, no first order effects exist.  

To compute orbits from Eq. (8),
we replace each $d x^{\mu}$ in the
generic form $ds^2 = g_{\mu\nu}dx^\mu dx^\nu$ by $dx^\mu/d\ta$ to
obtain the Lagrangian for motion along the $\theta = 0$ plane as
 \begin{equation} 2L=[1+2\Ph(\rb)]\lpar\fr{dt}{d\ta}\rpar^2-[1-2\Ph(\rb)]
 \left[\lpar\fr{d\rb}{d\ta}-H\rb\fr{dt}{d\ta}\rpar^2
 +\rb^2\lpar\fr{d\tha}{d\ta}\rpar^2\right], \end{equation}
where $\tau$ is the proper time, and
a factor of two was introduced on the left side by convention
(so that the canonical momenta $\partial L/\partial\dot t=\dot t$ and
$\partial L/\partial\dot r=a^2\dot r$, rather than twice these
quantities).  The three Euler--Lagrange (or geodesic) equations
corresponding to $t,\rb,\tha$ are, respectively:
 \begin{equation} \fr{d}{d\ta}[(1+2\Ph)\dot t+(1-2\Ph)H\rb(\dot\rb-H\rb\dot t)]=
 (1-2\Ph)\fr{dH}{dt}\rb\dot t(\dot\rb-H\rb\dot t),\end{equation}
 \begin{equation}-\fr{d}{d\ta}[(1-2\Ph)(\dot\rb-H\rb\dot t)]=
 \fr{d\Ph}{d\rb}[\dot t^2+(\dot\rb-H\rb\dot t)^2]
 +(1-2\Ph)[H\dot t(\dot\rb-H\rb\dot t)-\rb\dot\tha^2],\end{equation}
 \begin{equation} -\fr{d}{d\ta}[(1-2\Ph)\rb^2\dot\tha]=0, \end{equation}
where the `dot derivatives' are w.r.t. the proper time; e.g. $\dot t =
dt/d\tau$.
Since Eq. (9) is a homogeneous function of the time derivatives, there is an
immediate first integral: $L$ is a constant.  And to normalize $\ta$ to be
the proper time, we take $2L=1$.

We are interested in the effect of expansion on circular orbits
around a mass concentration, so
let us look for solutions in which (at least on
time scales short compared with 
the Hubble time) $\rb$ is a constant.  Moreover,
$\dot\tha$ and $\dot t$ are constants, the ratio between them
gives the angular
velocity:
 \begin{equation} \om=\fr{d\tha}{dt}=\fr{\dot\tha}{\dot t}. \end{equation}
Note that on the right side of Eq. (10) the quantity
$dH/dt$ is either multiplied by
$\dot\rb$, which we have assumed is zero, or by $H$.  Since
$dH/dt=-(1+q)H^2$ where $q = -a\ddot a/\dot a^2$ depicts the
(negative) acceleration of the expansion, the product $HdH/dt
 \sim H^3$ and can be ignored if we are only concerned with
terms of order $H^2$ or lower.

Under our assumptions, the left-hand sides of Eqs. (10) to 
(12)  all vanish, as do
the right-hand sides of Eqs. (10) and (12).  So we are left with one equation,
 \begin{equation} 0=\fr{d\Ph}{d\rb}(1-H^2 \rb^2)\dot t^2-(1-2\Ph)(H^2 \rb\dot t^2
 +\rb \dot\tha^2), \end{equation}
or, equivalently,
 \begin{equation} (1-2\Ph)(H^2+\om^2) \rb =(1-H^2 \rb^2)\fr{d\Ph}{d\rb}=
 (1-H^2 \rb^2)\fr{GM}{\rb^2} \end{equation}
where in the rightmost expression $M$ denotes the {\it excess} mass within
radius $\rb$, after the mass contribution to that region from the
mean density of matter and `dark energy' in the Universe is subtracted.
Solving for the angular velocity of the circular orbit, we obtain
 \begin{equation} \om^2=\fr{1-H^2 \rb^2}{1+\frac{2GM}{\rb}}
\fr{GM}{\rb^3}-H^2. \end{equation}
Provided that $GM/\rb \ll 1$ and $H^2 \rb^2 \ll 1$, Eq. (16) may be
recast as simply
 \begin{equation} \om^2= \fr{GM}{\rb^3}-H^2, \end{equation}
i.e. the net effect of the Hubble expansion is to reduce the angular velocity
w.r.t. its value under the scenario of pure
Newtonian gravity (or more precisely a Newtonianly
perturbed Minkowski line element).  A remarkable feature of Eq. (17) is 
that it predicts the existence of an {\it outermost bound orbit}, which has
a radius $R$ given by
\begin{equation} \fr{GM}{R^3} = H^2,~{\rm or}~\frac{v}{R} = H. \end{equation}
Eq. (18) yielded a radius smaller than that of Eq. (4).

It is also possible to define the outermost orbit in terms of an overdensity
criterion.  If in Eq. (18) we write $M = 4\pi R^3 \rho_{{\rm clump}}/3$,
where $\rho_{{\rm clump}}$ is the mean {\it over}density  within radius $R$,
the criterion will read
\begin{equation} \rho_{{\rm clump}} = \frac{3H^2}{4\pi G}. \end{equation}
For mass clumps at $z =$ 0 we have $H=H_0$, and the rightside of Eq. (19)
becomes twice the critical density, $2\rho_c$.  Thus we shall henceforth
refer to the boundary radius of a self-gravitating mass clump as
\begin{equation} r_2 = \left(\frac{GM}{H^2}\right)^{\frac{1}{3}}, \end{equation}
even if the terminology is obviously loose, in the sense that for
$z >$ 0 clumps we have $H > H_0$ and the rightside is no longer strictly
equal to $2\rho_c$.

\vspace{2mm}

\noindent
{\bf 3. Observational consequences: rotation
curves and the total matter budget}

Although the impression one gets from a
superficial perusal of the X-ray images of clusters of galaxies is that
clusters extend to radii of 1 -- 2 Mpc, commensurate with
the value of $r_{200}$ as defined in section 1, there are plenty of evidence
pointing to the existence of massive extended
halos in clusters and galaxies, i.e. the cutoff radius is more
consistent with $r_2$ of Eq. (20) than with $r_{200}$.  We provide examples in
each case to substantiate our claim.

\noindent
{\it 3.1 Clusters}

A typical number for the lower mass range of rich clusters is
$M =$ 10$^{15}$ M$_\odot$.  By Eq. (20) a $z =$ 0 cluster
has boundary radius
at $r_2 \approx$ 10~Mpc, whereas the same for a $z=$ 0.4 cluster is
$r_2 \approx$ 8.5 Mpc (we adopted $h =$ 0.7 throughout this subsection).
Thus, whether one `weighs' a cluster by
measuring the velocity dispersion of member galaxies, or 
assembling a weak lensing shear map from background sources, the
easier task is to target at higher redshifts: not only is a
cluster's  angular size smaller because of the larger distance, but
also its limiting physical radius $r_2$ is reduced.
Weak lensing observations did reveal a very
extended halo for the $z =$ 0.395 cluster
CL0024+1654 (Kneib et al 2003).  The cluster mass as quoted in this paper is
$M_{200} \approx$ 6 $\times$ 10$^{14}$ M$_\odot$, so that one
expects $r_2 >$ 7.2 Mpc (most likely much larger).  Indeed,
the data indicated a density profile $\rho_m(r) \sim 1/\rb^{2.4}$ 
continuing through the instrumental sensitivity limit at $\br \approx$ 5 Mpc
without any sign of cutoff.

An important test of the standard model can be conducted if rotation
(velocity dispersion) curves of clusters are measured out to
$\br = r_2$.  This means, by Eq. (20), that one needs data out to 10 Mpc
radii for nearby clusters, smaller for higher redshift clusters. 
Assuming the halo profile
at the outskirts has the asymptotic NFW form, then, by Eq. (17), the
observed circular velocity will scale with radius as
\begin{equation}
v_{{\rm obs}}^2 = \frac{G(M_0 + M_c {\rm ln} \rb)}{\rb} - H^2 \rb^2,
\end{equation}
where $M_c = 4\pi r_s^3 \rho_c \delta_c$ and $M_0 = M_{200} - 
M_c {\rm ln} r_{200}$.  Thus, while pure Newtonian gravity 
(the first term on the right side of the equation) predicts an
essentially $v_{{\rm obs}} \sim 1/\sqrt{\rb}$ decline for
$r_s \ll \rb \ll r_2$, for an expanding Universe
as $\rb \rightarrow r_2$ there is a sharp
dive towards $v_{{\rm obs}} =$ 0.  If the halo profile is an isothermal
sphere, where $\rho(\rb) \sim 1/\rb^2$, we will have a flat rotation
curve (i.e. $v_{{\rm obs}} = v_0$, a constant) when $H =$ 0.  With
expansion, however, $v_{{\rm obs}}^2 = v_0^2 - H^2 \rb^2$, again an
abrupt cutoff at $\br=r_2$.

For the Virgo cluster, the data for this test are either available
or (with databases like the SDSS) imminently so.  The fact that there is
controversy over how the Local Group relates to the cluster implies
galaxies as far out as 12-13 Mpc radii
are still  members of Virgo.  Note that at these radii
$H_0 r \rightarrow$ 1,000 km s$^{-1}$. i.e. the true velocities
(due to gravity alone) have to be well in excess of
$H_0 r$ in order to maintain the rotation curve against
rapid decline.   Thus, either Virgo has a total mass $\gg$ 10$^{15}$
M$_\odot$, or the predicted outermost orbit does not correspond 
to reality.
A closer look into this problem is definitely
priority task.

\noindent
{\it 3.2 Galaxies}

Weak lensing of background sources by foreground galaxies was
investigated by Hoekstra, Yee, \& Gladders (2004), and
Hoekstra et al (2005), who found clear
signals out to at least $\br=r_{200}$.  Based upon
the lensing data alone, no definitive statements
could be made about what lies beyond, apart from the fact that there
was no indication whatsoever of $r_{200}$ as representating any real
cutoff radius.  It is possible, nonetheless, to derive an average matter
density $\bar\rho_{\rm g}$
for galaxies from the Hoekstra et al observations,
and to investigate whether the inclusion of extra matter between
$r_{200}$ and $r_2$ would lead to a revised value for 
$\bar\rho_{\rm g}$ that is
closer to expectation.  

According to Hoekstra, Yee, \& Gladders (2004), the mass within 
$\br=r_{200} =$ 139$h^{-1}$ kpc is 
$M_{200} =$ 8.4 $\times$ 10$^{11}h^{-1}$ M$_\odot$
when the data for a representative galaxy were modelled with the NFW profile.
From Eq. (20), we see that even without any outlying matter, the $r_2$ radius
for $M=M_{200}$ is $r_2 \approx 1h^{-1}$ Mpc.  Since, in this
paper the scale radius of the NFW profile has the fitted value
of $r_s \approx$ 16$h^{-1}$ kpc, we have $\br \gg r_s$ for all
radii $r_{200} < \br < r_2$, so that the NFW profile reduces to
$\rho_{\rm g} (\br) = \delta_c \rho_c r_s^3/\br^3$,
which can readily be integrated from 
$\br=r_{200}$ to $\br=r_2$.  The outcome is,
of course, the mass of the remaining galactic matter halo, viz.
\begin{equation}
M(r_{200} \leq \br \leq r_2) = 4\pi r_s^3 \rho_c \delta_c~{\rm ln} \left(
\frac{r_2}{r_{200}}\right).
\end{equation}
By using the best-fit $\delta_c$ parameter of $\delta_c =$ 2.4 $\times$
10$^4$ as quoted in the paper, we then deduce that 
$M(r_{200} \leq \br \leq r_2)$ equals $\sim$ 85 \% of $M(\br <r_{200})$.  It
is clear also that in reality the halo boundary has radius
$>$ 1$h^{-1}$ Mpc because of the extra mass from the 
$r_2 > \br > r_{200}$ region
which we did not take into account when calculating $r_2$.  Thus we arrive 
at the comparison
\begin{equation}
M(\br < r_{200}) \approx M(r_{200} \leq \br \leq r_2)
\end{equation}
for NFW galaxy halo profiles, i.e. the inclusion of outlying matter would
usually lead to a {\it doubling of the total mass}.

In order to derive $\bar\rho_{\rm g}$, it is necessary to combine
galaxy mass and luminosity  measurements, and to obtain a luminosity
density for the same sample.  The former was done in Hoekstra et al
(2005), which still reported a representative value of $M_{200}$ close
to that quoted in the previous paragraph, and which also determined an
average B band  mass-to-light ratio of $M_{200}/L \approx$ 60 $h$ 
M$_\odot/$L$_\odot$ (see the bottom left plot of Figure 8 of Hoekstra et al
2005).  The latter is to be extracted from the CNOC2 survey
of Lin et al (1999), which targeted exactly the same range of galaxies
as those of Hoekstra et al (2005) and Hoekstra, Yee, \& Gladders (2004).
From Table 3 of Lin et al (1999), the total B band luminosity density
for 0.25 $< z <$ 0.4 galaxies (same redshift interval as that of the
Hoekstra et al 2005 lens sample) is $\rho_L^B =$ 1.2$h \times$ 10$^{20}$
W~Hz$^{-1}~$Mpc$^{-3}$.  Since the B band solar luminosity is $L_\odot^B =$
2.19 $\times$ 10$^{11}~$ W~Hz$^{-1}$, we may now couple this with the
aforementioned mass-to-light ratio to arrive at an average mass density of
\begin{equation}
\bar\rho_{\rm g} \approx 3.29 \times 10^{10} h^2~{\rm M}_\odot~{\rm Mpc}^{-3}.
\end{equation}
When comparing with the critical density of 6.11 $\times$ 10$^{11}h^2$
M$_\odot~$Mpc$^{-3}$
at $z=$ 0.3, the mean redshift of
the Hoekstra et al (2005) lenses, we see that the galaxies account for
5.4 \% of the matter at $z=$ 0.3.

Is this percentage reasonable?  From the
matter budget analysis of Fukugita (2004) and Fukugita, Hogan,
\& Peebles (1998) emerges the picture that $\sim$ 50 \% of the matter
in the near Universe is still `missing', and may well reside in
galaxies and their extended halos.  If $\Omega_m \approx$ 0.27,
then the expectation is
$\Omega_{\rm g} \approx$ 13.5 \%, i.e. $\sim$ twice the percentage value
as our 5.4 \%, which of course was inferred from the galaxy mass-to-light
ratio where mass refers to $M_{200}$.  Given however, that we
demonstrated a doubling of a galaxy's mass when the matter between
$\br=r_{200}$ and $\br=r_2$ is included (i.e. Eq. (22)),
the conclusion of a very extended halo
component of baryons and
dark matter fulfilling the anticipation of Fukugita (2004)
would appear in order.

\noindent
{\it 3.3 Groups}

Groups of galaxies are the `dark horse', in the sense that they
pose a major systematic uncertainty to the matter budget
of the near Universe, also in the direction of {\it raising}
the fraction of matter in halos.  The difference from clusters and
galaxies is that, while the former harbors negligible fraction of $\Omega_m$
and the latter $\sim$ 50 \% of $\Omega_m$
if halos are included, groups can potentially
be the refuge for a great deal more matter than either of them.  
Lieu \& Mittaz (2005) analyzed
the ESO survey database of 1,168 nearby 
groups (Ramella et al 2002) and found, at $h =$ 0.7,
a number density of 1.56
$\times$ 10$^{-4}$ Mpc$^{-3}$ and a mean
virial mass of $\bar M \approx$ 1.15 $\times$ 10$^{14}$ M$_\odot$ per group.
This leads, without inclusion of any extra mass that may be in
halo extensions, to a group matter density of $\Omega_m/2$.
Moreover, there is another effect.
In the ESO survey, the peak velocity dispersion
is $\sigma \approx$ 70 km~s$^{-1}$, with a corresponding virial mass of
$M \approx$ 10$^{12.5}$~M$_\odot$ (see Figure 1 of Lieu \& Mittaz 2005).
The radius at which $\sigma$ applies
is therefore at $\br \approx GM/(2\sigma^2) =$ 2.66 Mpc, so that $H_0 \br$
for $h =$ 0.7 is not far below the circular velocity 
$v = \sqrt{2} \sigma$, i.e.
we are again in a regime where, by Eq. (17), the observed velocity (hence
inferred mass) for these groups
is biased low by the Hubble effect on the orbits.  Thus,
however one looks at them, groups have a tendency to become `heavier',
which is why some authors 
(e.g. Guimaraes, Myers, \& Shanks 2005)
believe that the matter content of groups is a number to be
reckoned with.

In conclusion, the chief
emphasis of this paper concerns
the derivation of $r_2$, and the observational evidence for its
significance, including the
estimate that galaxies can account for the missing 50 \% of the
WMAP matter density at low redshifts if the halo mass between $r=r_{200}$
and $r=r_2$ is counted.  An important test of the standard model is
to search for the decline of cluster velocity dispersions near $\br = r_2$.
The model will be challenged if this decline is {\it not} seen
out to radii $r > r_2$, as seems to be the case for Virgo.

We thank T.W.B. Kibble, N. Gnedin, S. DeDeo,
and M. Chodorowski for helpful discussions, especially for
pointing out the absence of first order effects.

\vspace{1cm}

\noindent
{\bf References}



\noindent
Dubinski, J., \& Carlberg, R.G. 1991, ApJ, 378, 496.

\noindent
Eke, V.R., Cole, S., \& Frenk, C. 1996, MNRAS, 282, 263.


\noindent
Fukugita, M., 2004, in IAU Symp. 220, Dark matter in galaxies, ed. S.D.\\
\indent Ryder et al. (San Francisco: ASP), 227.

\noindent
Fukugita, M., Hogan, C.J., \& Peebles, P.J.E., 1998, ApJ, 503, 518.

\noindent
Futumase, T. \& Sasaki, M. 1989, Phys Rev D, 40, 2502.

\noindent
Guimaraes, A.C.C., Myers, A.D., \& Shanks, T. 2005, MNRAS, 362, 657.

\noindent
Hoekstra, H., Yee, H.K.C., \& Gladders, M.D., 2004, ApJ, 606, 67.

\noindent
Hoekstra, H., Hsieh, B.C., Yee, H.K.C., Lin, H., \& Gladders, M.D., 2005,
ApJ, 635, 73.

\noindent
Kneib, J. -P., Hudelot, P., Ellis, R.S., Treu, T., Smith, G.P., Marshall, P.,\\
\indent Czoske, O., Smail, I., \& Natarajan, P., 2003, ApJ, 598, 804.

\noindent
Lahav, O., Lilje, P. B., Primack, J. R., \& Rees, M. J., 1991, MNRAS, 251, 128.

\noindent
Lieu, R., \& Mittaz, J.P.D., 2005, ApJ, 628, 583.


\noindent
Lin, H., Yee, H. K. C. Carlberg, R. G., Morris, S. L., Sawicki, M., 
Patton, D.R.,\\
\indent Wirth, G., \&  Shepherd, C.W., 1999, ApJ, 518, 533.
	
\noindent
Lokas, E.L, \& Mamon, G.A. 2003, MNRAS, 343, 401.

\noindent
McVittie, G.C., 1933, MNRAS, 93, 325.

\noindent
Navarro, J.F., Frenk, C.S., \& White, S.D.M. 1995, MNRAS, 275, 56.

\noindent
Navarro, J.F., Frenk, C.S., \& White, S.D.M. 1996, ApJ, 462, 563.

\noindent
Navarro, J.F., Frenk, C.S., \& White, S.D.M. 1997, ApJ, 490, 493.


\noindent
Peebles, P.J.E., 1984, ApJ, 284, 439.

\noindent
Ramella, M., Geller, M.J., Pisani, A., \& da Costa, L.N.,
2002, AJ, 123, 2976.

\noindent
Ramella, M. et al, 1999, A \& A, 342, 1.

\noindent
Sandage, A. 1986, ApJ, 307, 1.

\noindent
Spergel, D., et al 2006, ApJ, submitted.

\end{document}